\documentclass[sigconf, nonacm]{acmart}

\newcommand\vldbdoi{XX.XX/XXX.XX}
\newcommand\vldbpages{XXX-XXX}
\newcommand\vldbvolume{14}
\newcommand\vldbissue{1}
\newcommand\vldbyear{2020}
\newcommand\vldbauthors{\authors}
\newcommand\vldbtitle{\shorttitle} 
\newcommand\vldbavailabilityurl{https://github.com/LJHzju/LASER}
\newcommand\vldbpagestyle{plain} 

\usepackage[caption=false,font=normalsize,labelfont=sf,textfont=sf]{subfig}
\usepackage[linesnumbered,vlined,ruled]{algorithm2e}
\usepackage{setspace}
\usepackage{enumitem}
\usepackage{balance}
\usepackage{graphicx}
\usepackage{caption}
\usepackage{subcaption}
\usepackage{multirow}
\usepackage{multicol}
\usepackage{colortbl}
\usepackage{makecell}
\usepackage{listings}
\usepackage{pifont}
\usepackage[a-2b]{pdfx}

\begin{document}
\title[LASER: A Data-Centric Method for Low-Cost and Efficient SQL Rewriting based on SQL-GRPO]
{LASER: A Data-Centric Method for Low-Cost and Efficient SQL Rewriting based on SQL-GRPO}

\author{Jiahui Li}
\affiliation{%
  \institution{Zhejiang University}
}
\email{li.jiahui@zju.edu.cn}

\author{Tongwang Wu}
\affiliation{%
  \institution{Zhejiang University}
}
\email{tongwang.wu@zju.edu.cn}

\author{Yuren Mao$^*$}
\affiliation{%
\institution{Zhejiang University}
}
\email{yuren.mao@zju.edu.cn}

\author{Rong Kang}
\affiliation{%
\institution{ByteDance Inc}
}
\email{kangrong.cn@bytedance.com}

\author{Tieying Zhang$^*$}
\affiliation{%
\institution{ByteDance Inc}
}
\email{tieying.zhang@bytedance.com}

\author{Yunjun Gao}
\affiliation{%
\institution{Zhejiang University}
}
\email{gaoyj@zju.edu.cn}

\begin{abstract}
Query rewriting, the process of transforming queries into semantically equivalent yet more efficient variants, is crucial for database optimization. Existing solutions predominantly rely on either rule-based heuristics or Large Language Models (LLMs). However, traditional rule-based methods lack adaptability, while LLM-based approaches incur prohibitive inference costs and privacy risks. In contrast, Small Language Models (SLMs) present a compelling middle ground, potentially offering both flexibility and efficiency. However, the development of such compact models is severely bottlenecked by the scarcity of high-quality, domain-specific training data.
To bridge this gap, we introduce LASER, a data-centric framework designed to empower small models for robust SQL optimization.
First, to address the scarcity of existing benchmarks and the limited optimization headroom of generic synthetic queries, we construct SQL-MCTS, a large-scale corpus of complex slow queries. We employ an MCTS-based hybrid expansion strategy that combines rule-guided anti-patterns with LLM mutations to evolve structurally expressive seeds into execution-verified slow variants.
Second, to enable the model to autonomously discover latency-aware rewriting patterns, we propose SQL-GRPO, a specialized alignment strategy adapted from Group Relative Policy Optimization. By integrating Anchored Group Advantage to refine advantage estimation and Complexity-Adaptive Dynamic Rollout to efficiently allocate exploration budgets, this approach effectively empowers compact models to master execution-based optimization logic.
Implemented on Qwen3 models, LASER significantly outperforms rule-based systems and LLMs in execution efficiency, while exhibiting robust zero-shot transferability with minimal overhead.
\end{abstract}

\thanks{$^*$Yuren Mao and Tieying Zhang is the corresponding author.}

\maketitle

\pagestyle{\vldbpagestyle}
\begingroup\small\noindent\raggedright\textbf{PVLDB Reference Format:}\\
\vldbauthors. \vldbtitle. PVLDB, \vldbvolume(\vldbissue): \vldbpages, \vldbyear.\\
\href{https://doi.org/\vldbdoi}{doi:\vldbdoi}
\endgroup
\begingroup
\renewcommand\thefootnote{}\footnote{\noindent
This work is licensed under the Creative Commons BY-NC-ND 4.0 International License. Visit \url{https://creativecommons.org/licenses/by-nc-nd/4.0/} to view a copy of this license. For any use beyond those covered by this license, obtain permission by emailing \href{mailto:info@vldb.org}{info@vldb.org}. Copyright is held by the owner/author(s). Publication rights licensed to the VLDB Endowment. \\
\raggedright Proceedings of the VLDB Endowment, Vol. \vldbvolume, No. \vldbissue\ %
ISSN 2150-8097. \\
\href{https://doi.org/\vldbdoi}{doi:\vldbdoi} \\
}\addtocounter{footnote}{-1}\endgroup

\ifdefempty{\vldbavailabilityurl}{}{
\vspace{.3cm}
\begingroup\small\noindent\raggedright\textbf{PVLDB Artifact Availability:}\\
The source code, data, and/or other artifacts have been made available at \url{\vldbavailabilityurl}.
\endgroup
}

\section{Introduction}
\label{sec:introduction}
With the rapid growth of data, SQL query optimization has become one of the key techniques for improving the performance of Database Management Systems (DBMS). In modern database applications, the efficiency of SQL queries directly impacts system response times and resource consumption. Therefore, SQL query rewriting has emerged as a crucial optimization method, widely employed to enhance query execution efficiency.

However, the task of rewriting queries is inherently challenging. The complexity arises from the need to transform a query into an equivalent form while improving its execution performance. Previous works broadly fall into two categories: (1) rule-based methods, which rely on predefined transformation rules to rewrite queries, and (2) LLM-based methods, which leverage large language models to generate new query formulations. However, existing methods face significant practical challenges, ranging from heavy data dependence to prohibitive costs and low execution efficiency.

\noindent
\textbf{Data Dependence and Poor Transferability.}
Several existing rewriting methods rely heavily on large quantities of data tailored to a specific database or benchmark. For example, LearnedRewrite~\cite{learnedrewrite} depends on a learned cost model trained on workload execution traces to guide its rule selection. Similarly, LLM-R$^2$~\cite{llm-r2} requires a substantial pool of SQL pairs for rule selection, while E$^3$-Rewrite~\cite{e3} trains its rewrite model on a large corpus of SQL queries. In practice, these datasets are typically synthesized from template-based benchmarks such as TPC-H~\cite{tpch}, and DSB~\cite{dsb}. The resulting workloads exhibit limited structural and semantic diversity, and the overlap between training and evaluation templates creates a high risk of data leakage. Models tuned in this way tend to overfit benchmark-style patterns, struggling to transfer to unseen schemas or real-world queries~\cite{overfit}.

\noindent
\textbf{High Cost and Privacy Risks.}
Most prior systems depend on proprietary, large-scale LLM APIs in the rewrite loop. LLM-R$^2$~\cite{llm-r2} and R-Bot~\cite{r-bot} invoke closed-source models to select or rank rewrite rules, whereas QUITE~\cite{quite} and GenRewrite~\cite{genrewrite} directly call such APIs to generate rewritten queries. This design incurs substantial monetary cost when serving many queries, since each rewrite may require multiple long-prompt API calls. More importantly, shipping production SQL, schema information, and sometimes execution feedback to external services raises serious privacy and compliance concerns, making these approaches difficult to deploy in settings with strict data governance or regulatory constraints.

\noindent
\textbf{Low Rewriting Efficiency.}
LLM-based methods also tend to have low end-to-end rewrite efficiency. Systems such as R-Bot~\cite{r-bot} perform multiple rounds of LLM inference to explore candidate rule sequences, and then rely on engines like Apache Calcite~\cite{calcite} to apply the selected rules, resulting in substantial per-query overhead. QUITE~\cite{quite} adopts a multi-step, agent-style workflow in which LLM agents iteratively analyze, plan, and refine query rewrites. Across these designs, repeatedly querying large models, orchestrating multi-stage tool calls, and executing validation passes can make the latency of a single rewrite orders of magnitude higher than that of native optimizer decisions. This low efficiency becomes a critical bottleneck for real-time or interactive scenarios.

Given these limitations, specialized Small Language Models offer a compelling solution to the trilemma of adaptability, efficiency, and privacy. Local deployment mitigates the latency and privacy risks of LLMs API-dependent workflows while maintaining reasoning capabilities. Crucially, Chinchilla Scaling Laws~\cite{chinchilla} suggest that performance depends more on data quality than on parameter count alone. This implies that compact models can achieve expert-level SQL optimization when aligned with high-quality, domain-specific data.
However, realizing this potential faces two primary hurdles. First, current training data resources are bottlenecked by the scarcity of public benchmarks and the limited optimization headroom of generic synthetic queries, which lack the structural complexity to capture the causal mapping between rewriting and latency reduction. Second, Reinforcement Learning (RL) is theoretically ideal for optimization tasks by leveraging execution feedback~\cite{rlsft, neo, balsa}. In particular, Group Relative Policy Optimization offers a compelling low-cost avenue for policy alignment, having been widely adopted in mathematical and reasoning domains~\cite{grpo, reasoning-sql,  posterior-grpo}. However, directly applying standard GRPO to SQL rewriting proves ineffective, as it typically suffers from misleading advantage estimation caused by invalid candidates and inefficient exploration budget allocation, preventing the model from autonomously mastering latency-aware patterns.

To overcome these challenges, we propose LASER, a data-centric method for Low-cost and Efficient SQL Rewriting. LASER implements a unified framework that synergizes high-quality data synthesis with policy optimization.
In the first stage, to address the scarcity and triviality of existing data, we construct SQL-MCTS. Starting from structurally expressive seed queries, the framework employs an MCTS-based hybrid expansion strategy that combines rule-guided anti-patterns with LLM mutations to evolve these seeds into execution-verified slow variants, thereby capturing the causal mapping between structural degradation and latency to provide necessary optimization headroom.
In the second stage, we perform SQL-GRPO, a specialized alignment framework. Specifically, this strategy incorporates Anchored Group Advantage to rectify false-positive signals by grounding relative rewards against absolute baselines, and employs Complexity-Adaptive Dynamic Rollout to dynamically concentrate exploration resources on queries with higher structural difficulty. By combining this with an SFT cold-start and Verification-Driven Self-Correction, LASER yields a compact model that achieves strong optimization performance at low inference cost and can be transferred across databases.

Our contributions are summarized as follows:
\begin{itemize}
    \setlength\leftskip{-2em}
    \item We propose LASER, a data-centric framework for low-cost and efficient SQL rewriting that synergizes automated slow query generation with a specialized training pipeline to empower compact models with robust optimization capabilities (Section~\ref{sec:overview}). 
    \item We construct SQL-MCTS, a slow-query dataset on the TPC-DS schema containing 11,675 generated queries. These queries exhibit superior structural complexity compared to existing benchmarks, providing a rich and challenging foundation for training SQL rewrite models (Section~\ref{sec:generation}).
    \item We develop a specialized optimization pipeline centered on SQL-GRPO, a novel alignment strategy that incorporates Anchored Group Advantage and Complexity-Adaptive Dynamic Rollout. This enables compact models to autonomously master execution-based optimization logic beyond static supervision (Section~\ref{sec:training}).
    \item Extensive experiments demonstrate that LASER achieves superior rewrite performance across diverse benchmarks, while exhibiting robust zero-shot transferability. Crucially, it achieves these results locally with minimal inference latency, validating its suitability for privacy-sensitive and resource-constrained deployments (Section~\ref{sec:experiments}).
\end{itemize}

\section{Related Work}
\label{sec:relatedword}

\subsection{Query Rewrite}
SQL query rewrite refers to the process of transforming a given SQL query into an equivalent form that executes more efficiently. Query rewrite is a critical step in modern Database Management Systems (DBMSs), as it can significantly improve the performance of complex queries without altering the underlying semantics. Existing query rewrite techniques can be broadly divided into two categories: rule-based query rewrite and LLM-based query rewrite.

\subsubsection{Rule-based Query Rewrite}
Rule-based query rewrite methods~\cite{wetune, learnedrewrite, llm-r2, r-bot} rely on a set of predefined transformation rules (e.g., predicate pushdown~\cite{predicatepushdown} and elimination of redundant operators) to improve execution efficiency. These rules are often implemented in an algebraic framework such as Apache Calcite~\cite{calcite} and are repeatedly applied until no further improvements are detected.

For example, WeTune~\cite{wetune} automatically generates and verifies logical plan transformations with a Cost-Based Optimizer (CBO), exploring alternative plans by composing algebraic equivalences. However, its search space is constrained to specific operator types and equivalence patterns, which limits its ability to capture richer rewrite opportunities. LearnedRewrite~\cite{learnedrewrite} combines Monte Carlo Tree Search (MCTS) with a learned cost model to explore rule applications, but its effectiveness heavily depends on the accuracy of the cost model.
More recently, LLM-R$^2$~\cite{llm-r2} and R-Bot~\cite{r-bot} integrate large language models into rule-based frameworks while still executing concrete transformations through predefined rules. LLM-R$^2$ uses an LLM to guide rule selection, while R-Bot retrieves structural and semantic evidence from historical queries to assist in rule ordering and application.

Despite their advances, these methods remain constrained by the expressiveness of the rule set. The rules are typically designed to cover common query patterns and standard relational operators. As a result, they are difficult to generalize to novel or highly complex query structures, such as intricate Common Table Expressions (CTEs) and deeply nested subqueries.

\subsubsection{LLM-based Query Rewrite}
LLM-based query rewrite methods~\cite{genrewrite, quite, e3} leverage the capabilities of LLMs to directly rewrite SQL queries. 
For example, GenRewrite~\cite{genrewrite} introduces Natural Language Rewrite Rules (NLR2s) to provide a textual explanation of the rewrite process, and employs an iterative process to rewrite and correct queries based on feedback loops. However, this process can be computationally expensive and requires multiple rounds of rewriting and validation to ensure correctness and efficiency.
QUITE~\cite{quite} enhances LLM-based query rewriting by introducing a multi-agent framework controlled by a finite state machine (FSM). However, the system's reliance on a multi-agent architecture and closed-source APIs can introduce scalability challenges, limiting its applicability in certain environments.
E$^3$-Rewrite~\cite{e3} utilizes Reinforcement Learning (RL) with open-source models to optimize queries for executability, equivalence, and efficiency. But due to the absence of a dedicated training dataset, it constructs training data from templates of the evaluation benchmarks, thus limiting its ability to generalize to other novel or complex queries~\cite{overfit}.

\subsection{SQL Generation} 
Effective training for SQL query rewriting requires diverse and high-quality datasets, but generating such datasets remains a significant challenge due to the lack of specialized training data. While benchmark datasets such as TPC-DS~\cite{tpcds} and TPC-H~\cite{tpch} offer predefined templates for SQL queries, these datasets fall short in terms of diversity and complexity, which limits their utility in training models for real-world applications.

To address these shortcomings, alternative methods such as OmniSQL~\cite{omnisql} and SQL-Factory~\cite{sql-factory} have been developed to automatically synthesize large-scale SQL queries. These approaches enhance query diversity by leveraging schema-aware generation techniques and more flexible query construction methods.
However, despite the improvements in query variety, they still face limitations in capturing the full structural complexity required for comprehensive query rewriting tasks. In practice, a considerable portion of the generated queries exhibit shallow join depth, few predicates, and limited use of nested subqueries or advanced operators, leaving little room for meaningful logical or physical rewrites. Moreover, these generators are primarily designed for workload coverage and diversity rather than for systematically constructing suboptimal or slow queries. As a result, many synthesized queries already resemble well-structured, near-optimal workloads and thus lack sufficient optimization headroom for training rewrite models.

\subsection{Reinforcement Learning in LLMs}
In domains such as query optimization and program synthesis, tasks often do not admit a single canonical ground-truth answer. For these tasks, supervision from paired data is either unavailable or too coarse to capture fine-grained preferences over alternative outputs, such as latency, resource consumption, or readability~\cite{balsa, sftrl1, sftrl2}. Consequently, Reinforcement Learning has become a natural complement to supervised fine-tuning, enabling models to be optimized directly for scalar feedback signals that reflect diverse task-specific quality criteria.

A prominent line of work employs RL to align LLMs with human or automated preferences, often referred to as Reinforcement Learning from Human Feedback (RLHF)~\cite{rlhf} and its variants~\cite{dpo, rloo, rto}. These methods typically rely on Actor-Critic algorithms~\cite{actor-critic} to update the policy based on rewards derived from preference models. However, this paradigm incurs significant memory and computational overhead due to the necessity of maintaining a separate value network (Critic) alongside the policy model. This high resource cost becomes a critical bottleneck when deploying efficient, low-latency training pipelines for specialized tasks.

To mitigate these costs, recent work introduces Group Relative Policy Optimization~\cite{grpo}, which eliminates the critic model by estimating baselines from the group-wise mean of generated outputs. While GRPO significantly improves training efficiency, its standard formulation assumes uniform sample difficulty and relies on relative scoring within a group. This poses unique challenges for SQL optimization, where workload heterogeneity leads to resource misallocation, and relative advantages can erroneously reward slow queries simply for being valid amidst a group of syntax errors.
\section{System Overview}
\label{sec:overview}

\begin{figure*}[!t]
    \centering
    \includegraphics[width=0.993\linewidth]{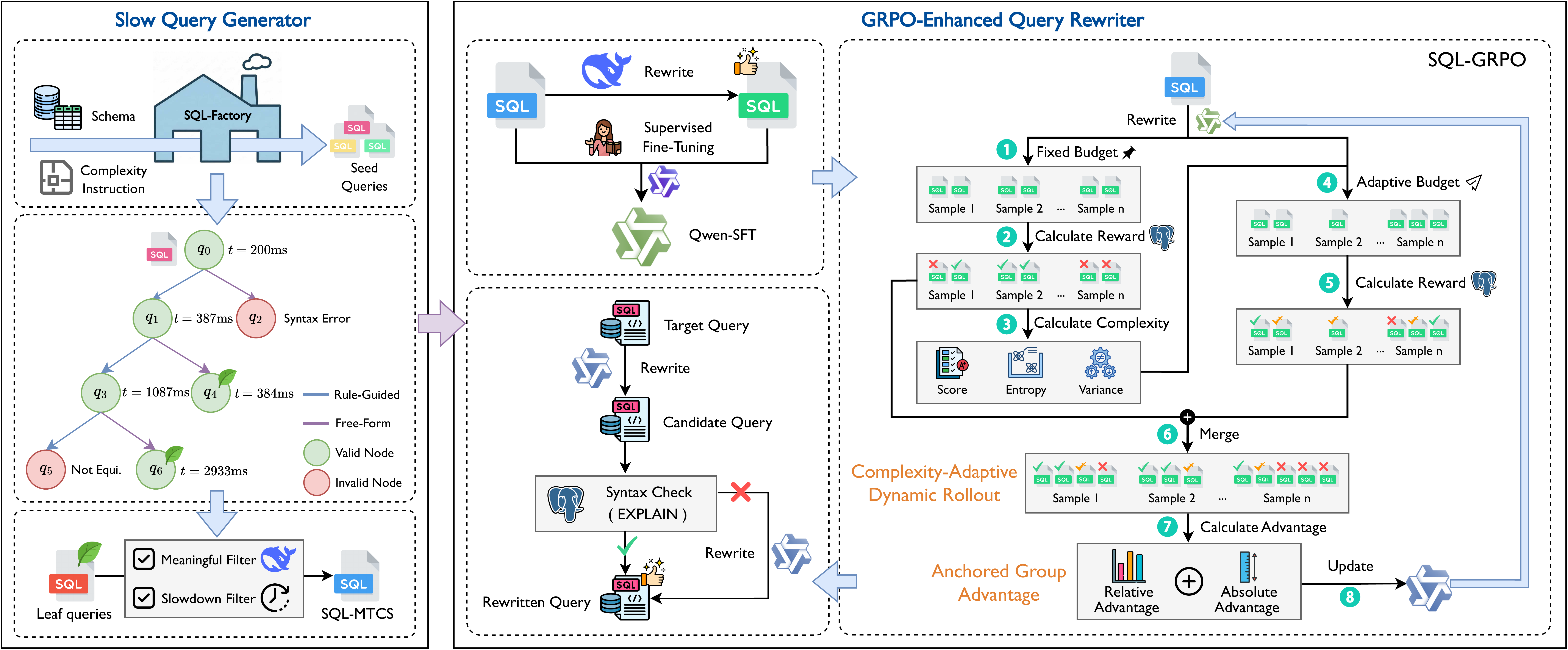} 
    \vspace{-2.5mm}
    \caption{An overview of LASER framework.} 
    \label{fig:overview}
    \vspace{-3mm}
\end{figure*}

This section presents LASER, a data-centric method for Low-cost and Efficient SQL Rewriting that enables lightweight models to perform effective SQL optimization. As illustrated in Figure~\ref{fig:overview}, LASER comprises two synergistic modules: (1) a \textit{Slow Query Generator} that constructs the SQL-MCTS dataset via MCTS-based evolution of complexity-aware seed queries, and (2) a \textit{GRPO-Enhanced Query Rewriter} that leverages our specialized SQL-GRPO algorithm to train the model to produce execution-efficient SQL rewrites.

\subsection{Slow Query Generator}

To support training and evaluation of our data-centric SQL rewriting method, we design a comprehensive framework for generating semantically equivalent but progressively slower SQL queries. While existing SQL generation systems (e.g., SQL-Factory~\cite{sql-factory}, OmniSQL~\cite{omnisql}) provide broad syntactic diversity, their outputs often lack the structural richness required for controlled performance degradation. Queries with shallow join structures or simplistic predicates offer limited opportunities for cost-increasing transformations, causing search-based rewriting approaches to stagnate prematurely. This limitation motivates the development of a specialized slow query generation module capable of producing structurally expressive seeds and guiding their evolution toward performance-inefficient variants.

Our framework, as shown in the left panel of Figure~\ref{fig:overview}, is composed of two major components: Complexity-Aware Seed Initialization and MCTS-Driven Slow Query Generation. Each component plays a distinct and complementary role in constructing SQL-MCTS, a slow-query corpus specifically tailored for model training.

\noindent
\textbf{Complexity-Aware Seed Initialization.} This component focuses on producing structurally expressive seed queries that serve as the foundation for subsequent slow-down exploration. By injecting rigorous constraints into the SQL-Factory~\cite{sql-factory} framework, we explicitly steer the synthesis toward queries featuring high-density predicates, deep nesting, and multi-table joins. This ensures that the generated seeds possess sufficient relational interdependence to support extensive cost-increasing transformations. Furthermore, an execution-based validation layer filters out invalid or empty-result candidates, guaranteeing that the subsequent MCTS exploration begins with a robust and logically meaningful foundation.

\noindent 
\textbf{MCTS-Driven Slow Query Generation.} This component transforms seed queries into slower yet semantically equivalent variants. It employs a Monte Carlo Tree Search framework augmented with a hybrid expansion strategy and execution-based evaluation. To synergize the targeted exploitation of known optimizer weaknesses with the open-ended exploration of novel structures, the search process expands nodes through either rule-guided transformations derived from a library of anti-patterns based on Calcite, or free-form mutations produced directly by an LLM. Furthermore, all generated variants are executed to verify semantic correctness and benchmark performance, using this feedback to directly drive reward computation. Consequently, the MCTS procedure refines its policy to progressively discover severe performance regressions. To ensure the integrity of the final corpus, the raw variants undergo a rigorous consolidation phase involving latency-based filtering and model-based validity audits. The final collection of samples resulting from this pipeline constitutes the SQL-MCTS dataset.

\subsection{GRPO-Enhanced Query Rewriter}

Following the construction of the slow query dataset, the core objective of our system is to train a specialized rewrite model capable of autonomously discovering high-performance SQL variants. While large reasoning models like DeepSeek-R1 demonstrate general-purpose coding proficiency, deploying them directly for latency-sensitive optimization is hindered by prohibitive inference costs and a disconnect from physical execution contexts. This necessitates a coarse-to-fine training paradigm, where we first distill semantic optimization knowledge into a compact local model and subsequently align its policy with concrete execution feedback.

As depicted in the right panel of Figure~\ref{fig:overview}, our training pipeline comprises two sequential phases:

\noindent
\textbf{Supervised Fine-Tuning}. This phase addresses the cold-start problem inherent in reinforcement learning, where training from scratch is sample-inefficient and unstable. To mitigate this, we employ Supervised Fine-Tuning to distill the expert reasoning capabilities of DeepSeek-R1 into our target model. Specifically, we prompt the teacher model with the generated slow queries to autonomously synthesize dual-output trajectories comprising the optimized SQL rewrites and their corresponding Chain-of-Thought (CoT)~\cite{cot} rationales. These rationales explicate the underlying optimization logic, including techniques such as subquery decorrelation~\cite{decorrelation} and predicate pushdown~\cite{predicatepushdown}. Fine-tuning on these reasoning-augmented samples equips the model with a foundational understanding of SQL equivalence and optimization heuristics, establishing a robust initial policy for subsequent exploration~\cite{hes-sql}.

\noindent
\textbf{SQL-GRPO.} This phase transforms the initialized policy from a static imitator into an active explorer. We leverage Group Relative Policy Optimization to optimize directly for execution latency and correctness. However, standard GRPO struggles with the heterogeneity of SQL workloads, often leading to resource misallocation~\cite{greso, xrpo} and inappropriate advantage estimation. To address these limitations, we introduce SQL-GRPO, a specialized alignment framework featuring two novel mechanisms:

(1) \textit{Complexity-Adaptive Dynamic Rollout.} Recognizing the extreme heterogeneity in SQL rewriting, this mechanism surpasses uniform sampling by dynamically reallocating the rollout budget. It prioritizes instances where the model exhibits high uncertainty or struggles to bypass validity filters, ensuring sufficient exploration to discover sparse valid rewrites for complex queries while eliminating computational redundancy on early-saturating instances.

(2) \textit{Anchored Group Advantage.} Standard GRPO computes advantages relative to the group mean, which creates a critical flaw when applied to SQL generation. Specifically, a slow and mediocre rewrite can receive an inflated positive advantage simply by outperforming a batch of invalid candidates that failed syntax checks. To rectify this, we propose an anchored advantage estimator that incorporates an absolute performance baseline into the advantage computation. This mechanism effectively suppresses such false positive signals, ensuring that high advantages are assigned only to rewrites that achieve genuine latency reduction rather than those that merely avoid execution failure.

Through this synergistic approach, our system integrates the broad semantic reasoning distilled from large models with the precise, feedback-driven optimization enabled by our SQL-GRPO framework. Additionally, a verification-driven self-correction mechanism serves as the final safeguard for executability, collectively achieving robust and reliable performance gains in SQL rewriting.

\begin{figure*}[!t]
    \centering
    \includegraphics[width=0.996\linewidth]{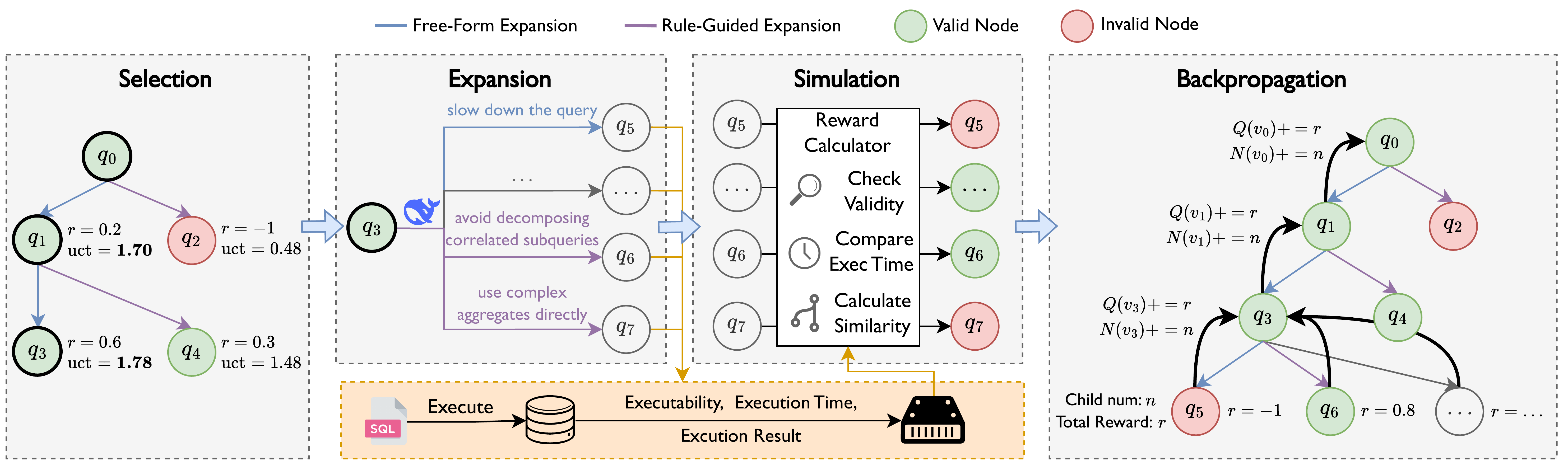} 
    \vspace{-2mm}
    \caption{The workflow of MCTS-Driven Slow Query Generation} 
    \label{fig:mcts}
    \vspace{-3mm}
\end{figure*}

\section{Slow Query Generator}
\label{sec:generation}

In this section, we detail the methodology for constructing SQL-MCTS, a large-scale corpus of high-quality slow queries that serves as the foundation for training our query rewriter. As illustrated in Figure~\ref{fig:overview}, the generation pipeline comprises two components: (1) Complexity-Aware Seed Initialization, which synthesizes structurally expressive seeds to ensure sufficient optimization headroom, and (2) MCTS-Driven Slow Query Generation, which systematically evolves these seeds into progressively slower variants via controlled performance degradation. Finally, we describe the dataset construction process to produce the final SQL-MCTS corpus.

\subsection{Complexity-Aware Seed Initialization}

A fundamental bottleneck in constructing a robust slow-query dataset is the structural simplicity of queries produced by generic generators like SQL-Factory~\cite{sql-factory} and OmniSQL~\cite{omnisql}. Default configurations often yield queries with narrow projections and sparse predicates, limiting the search space for subsequent cost-increasing mutations. To overcome this, we intervene in the \textit{Generation Team} component of the SQL-Factory framework by injecting rigorous, complexity-driven constraints directly into its generation prompt.

Specifically, these constraints mandate the synthesis of queries characterized by high-density predicate logic and relational interdependence. Departing from open-ended generation that favors simplistic patterns, the generator is explicitly steered to construct deeply nested subqueries and multi-table joins filtered by diverse logical operators. This ensures that the initial seeds possess sufficient structural complexity to support extensive rewriting, acting as robust anchors that allow the subsequent MCTS process to explore a non-trivial space of performance-degrading transformations.

Finally, to guarantee the applicability of these seeds, we employ an execution-based validation step to filter out invalid queries or those yielding empty result sets, ensuring that the initial corpus consists solely of executable and logically meaningful SQL statements.

\subsection{MCTS-Driven Slow Query Generation}

Given the complexity-aware seed pool, we employ a Monte Carlo Tree Search framework to systematically evolve each seed query into semantically equivalent but increasingly slower variants. Each SQL instance is treated as a node in the search tree, with edges representing cost-increasing transformations. Our procedure follows the classical four-phase MCTS~\cite{mcts-survy} loop consisting of Selection, Expansion, Simulation, and Backpropagation, as visually detailed in Figure~\ref{fig:mcts}.

\noindent\textbf{Selection.}
Starting from the root seed query, the algorithm recursively traverses the tree to select the most promising child node. At each step, given a parent node $v$, we choose the child $v_i$ from the set of valid children $\mathcal{C}(v)$ that maximizes the Upper Confidence Bound for Tree (UCT)~\cite{uct} score:
\begin{equation}
    v^* = \underset{v_i \in \mathcal{C}(v)}{\arg\max} \left( \frac{Q(v_i)}{N(v_i)} + c\sqrt{\frac{2\ln N(v)}{N(v_i)}} \right),
\end{equation}
where $Q(\cdot)$ is the cumulative reward, $N(\cdot)$ is the visit count, and $c$ is the exploration constant. The first term prioritizes exploitation by favoring high-reward nodes, while the second term drives exploration by probing less-visited branches. Selection continues until a leaf node is reached.

\noindent\textbf{Expansion.}
The algorithm expands each selected leaf node into multiple child nodes. For each new node, the system randomly selects a transformation strategy, applying either a targeted rule-based pattern or a free-form mutation.

(1) \textit{Rule-Guided Expansion.}
In this mode, we maintain a curated library of \textit{Slowdown Causes}, extracted by an LLM from Calcite rewrite rules by abstracting their inverse performance implications. Conceptually, this process operates as a Reverse Query Optimizer. While traditional optimizers (e.g., Calcite) utilize equivalence rules to transform inefficient patterns into efficient joins, our approach inverts this logic to systematically identify the anti-patterns that optimizers aim to eliminate. For instance, we explicitly invert decorrelation logic, transforming standard Equi-Joins into correlated subqueries to impede set-oriented processing.  This ensures the generated slowness is structurally inherent and optimization-solvable. To execute this, we employ a history-aware filter that selects only unused strategies along the current path, converting each chosen strategy into a structural prompt for the LLM.

(2) \textit{Free-Form Expansion.}  
Alternatively, to capture long-tail performance anti-patterns outside the defined rules, the LLM is also allowed to freely introduce inefficiencies such as unnecessary nesting, redundant filters, or deeply wrapped subqueries. This mode increases transformation diversity beyond what rule-based approaches can explicitly enumerate.

\noindent\textbf{Execution Dispatch.}
\label{sec:parallel}
To facilitate the ground-truth evaluation required by our framework, each generated variant is immediately submitted for asynchronous execution against a representative database instance. We utilize standard benchmark generation tools to create a scaled-down database that preserves the original schema and data distributions to balance evaluation fidelity with runtime efficiency. This non-blocking dispatch strategy pre-populates the result cache, ensuring that the subsequent Simulation phase can retrieve execution latency and correctness metrics without stalling the search pipeline.

\noindent\textbf{Simulation.}
Unlike standard MCTS formulations that rely on stochastic rollouts, our framework evaluates query utility via concrete execution on the target DBMS. Leveraging the asynchronous dispatch, this phase simply retrieves the finalized latency and correctness metrics from the result cache.

Upon retrieval, we assess the quality of node $v$ relative to the seed node $v_0$ based on execution latency $T_v$ and the result-set hash $H(v)$. To ensure deterministic verification, we define $H(\cdot)$ as the aggregate of row-level hash values computed after a mandatory in-memory sort of the result set. This sorting phase eliminates variations caused by non-deterministic row ordering under parallel execution, ensuring that the hash comparison reflects only semantic consistency of the result. Consequently, we define the query validity of node $v$ via the indicator function $\Phi(v)$:
\begin{equation}
\Phi(v) = \mathbb{I}\big[ \texttt{ExecSuccess}(q_v) \big] \cdot \mathbb{I}\big[ H(q_v) = H_0 \big],
\end{equation}
where $H(q_v) = H_0$ and $\texttt{ExecSuccess}(\cdot)$ indicates successful execution without runtime errors (excluding timeouts).

To quantify structural novelty, we assess the AST-level divergence. Let $\text{AST}(q)$ denote the Abstract Syntax Tree (AST) of query $q$, and $|\text{AST}(q)|$ represent the total number of nodes in the tree. We define the normalized structural distance $\Delta(q, q')$ as:
\begin{equation}
    \Delta(q, q') = \frac{\text{TED}(\text{AST}(q), \text{AST}(q'))}{\max(|\text{AST}(q)|, |\text{AST}(q')|)},
\end{equation}
where $\text{TED}(\cdot)$ measures the tree edit distance using SQLGlot~\cite{sqlglot}. The structural score $R_s(v)$ averages the divergence from both the parent $v_p$ and the seed $v_0$, defined as:
\begin{equation}
    R_s(v) = \frac{1}{2} \left[ \Delta(q_v, q_{v_p}) + \Delta(q_v, q_{v_0}) \right].
\end{equation}

Finally, we unify the reward assignment into a piecewise function to robustly handle execution failures, timeouts, and valid performance degradations. Specifically, let $T_{\mathrm{max}}$ be the DBMS timeout limit, the total reward $R(v)$ is defined as:
\begin{equation}
    \label{eq:mcts-reward}
    R(v) = 
    \begin{cases}
        \rho, 
        & \text{if } \Phi(v) = 0, \\
        
        \gamma, 
        & \text{if } T_v \geq T_{\mathrm{max}}, \\
        
        \lambda_{t} \tanh\!\left(\alpha \log \frac{T_v}{T_0}\right) + \lambda_{s} R_s(v), 
        & \text{otherwise},
    \end{cases}
\end{equation}
where $\rho$ penalizes invalid states. Crucially, since result equivalence cannot be verified for timed-out queries, we assign a static saturation reward $\gamma$. We set $\gamma = \lambda_t$ to acknowledge their high latency potential while withholding the structural bonus $\lambda_s$ as a penalty for semantic uncertainty. This configuration encourages the exploration of performance boundaries while preventing the search from converging solely on unverifiable variants. For verifiable executions, the reward combines the logarithmic slowdown factor and structural novelty, balanced by coefficients $\lambda_{t}$ and $\lambda_{s}$.

\noindent\textbf{Backpropagation.}
The reward obtained from simulation is propagated upward from the evaluated node to the root. Each node along the path updates its visit count and cumulative score via standard MCTS rules. High-reward transformations thus gain more influence on future traversal decisions, progressively steering the search toward regions of the space that yield substantial slowdowns.

Through repeated execution of these phases, the MCTS procedure converges toward structurally diverse, semantically preserved, and heavily degraded SQL variants, forming the core of our slow-query dataset.

\subsection{Dataset Construction}

Following the generation of candidate queries via MCTS, we consolidate the valid leaf nodes into a unified training corpus termed SQL-MCTS. This construction process filters and refines the raw search results through two rigorous stages: Latency-Based Filtering and Structural Validity Audit.

\noindent
\textbf{Identification of Slow Queries.}
From the leaf nodes of the MCTS tree, we strictly retain queries whose execution time is at least twice that of the original query. We classify these as slow queries. This filtering criterion effectively eliminates trivial modifications that create negligible performance variance, ensuring that every sample in SQL-MCTS represents a genuine performance deterioration suitable for learning optimization logic.

\noindent
\textbf{Slow Query Validity Check.}
To further guarantee the quality and relevance of the identified slow queries, we employ the DeepSeek-V3 model to audit the validity of the rewrites. The objective is to verify that the performance degradation stems from structural complexity, such as increased logical depth, rather than syntactic noise. This step effectively removes low-quality queries containing redundant or artificial operations, such as superfluous \texttt{ORDER BY} clauses or unnecessary subqueries, which contribute to latency without adding logical value.

Through this pipeline, we constructed the final SQL-MCTS dataset, comprising 11,675 high-quality slow queries evolved from 3,000 initial seeds on the TPC-DS~\cite{tpcds} benchmark schema. The entire construction process spanned approximately 6.3 days.

\section{GRPO-Enhanced Query Rewriter}
\label{sec:training}
In this section, we present the methodology for our GRPO-Enhanced Query Rewriting framework. Our approach follows a progressive three-stage pipeline: initializing the model via Supervised Fine-Tuning, optimizing execution performance using our specialized SQL-GRPO algorithm, and finally ensuring executability through an Inference with Verification-Driven Self-Correction mechanism.

\subsection{Supervised Fine-Tuning}
We initiate the training pipeline with Supervised Fine-Tuning to establish the model's foundational reasoning capabilities and SQL synthesis proficiency. This phase is crucial for aligning the policy with syntactic constraints and logical structures, providing a warm-start for the subsequent RL stage.

To construct the SFT corpus, we sample 30\% of the instances from our SQL-MCTS dataset, and leverage the reasoning capabilities of DeepSeek-R1~\cite{deepseek-r1} to distill expert-level optimization trajectories. The training inputs comprise the database schema, the original slow query, and its execution plan (retrieved via \texttt{EXPLAIN}), while the targets represent the reasoning trace and the optimized SQL output.
We fine-tune the Qwen3 model using the standard next-token prediction objective~\cite{sft}:
\begin{equation}
    \mathcal{L}_{SFT}(\theta) = -\frac{1}{T} \sum_{t=1}^{T} \log p_{\theta}(y_t | x, y_{<t}), 
\end{equation}
where $\theta$ denotes the model parameters, $x$ and $y$ represent the input and corresponding output sequences. Specifically, $y_t$ is the $t$-th token in the output, and $y_{<t}$ represents the tokens predicted before time step $t$. This phase equips the model with essential structural knowledge, ensuring that the initial policy generates logically valid SQL queries before performance optimization begins.

\subsection{SQL-GRPO}
While SFT ensures syntactic correctness, it primarily imitates reasoning patterns rather than directly optimizing for latency. To explicitly align the model with performance objectives, we employ Group Relative Policy Optimization~\cite{grpo}. By estimating baselines from the mean reward of group-wise outputs, GRPO efficiently eliminates the need for a separate value network. However, standard GRPO assumes uniform sample difficulty and scale-invariant advantages. These assumptions fail in SQL rewriting, which is characterized by strict validity constraints and significant performance variance. To address these challenges, we propose SQL-GRPO, which introduces a Performance-Driven Hierarchical Reward function alongside two architectural modifications: Complexity-Adaptive Dynamic Rollout and Anchored Group Advantage.

\noindent
\textbf{Performance-Driven Hierarchical Reward.}
Unlike general text generation tasks, SQL rewriting is governed by strict validity constraints where syntactically invalid SQL queries yield zero functional utility. To address this, we design a piecewise, hierarchical reward function that establishes an implicit curriculum. This structure encourages the model to prioritize the mastery of syntactic correctness and semantic equivalence before attempting the more challenging objective of latency optimization.

Crucially, we derive the reward signal from physical execution latency rather than theoretical optimizer cost estimates, which are prone to significant cardinality estimation errors in complex queries~\cite{base}. To mitigate runtime overhead, we adopt the asynchronous execution strategy utilized in the MCTS phase (Section~\ref{sec:parallel}). This parallelization allows database execution to proceed concurrently with policy rollout, significantly offsetting the execution latency against the generation time of subsequent samples.

Let $q_0$ be the seed query with baseline latency $T_0$, and $q'$ be the SQL rewrite candidate extracted from the model response. The reward function $R(q')$ imposes a strict validity hierarchy:
\begin{equation}
    R(q') = 
    \begin{cases}
        \rho_{\text{fmt}}, & \text{if } \texttt{ExtractionFail}(q'), \\
        \rho_{\text{exe}}, & \text{if } \texttt{ExecError}(q'), \\
        \rho_{\text{sem}}, & \text{if } \neg \texttt{Equiv}(q', q_0), \\
        \mathcal{F}(T_{q'}, T_0), & \text{if } \texttt{Success}(q').
    \end{cases}
\end{equation}
We set $\rho_{\text{fmt}} < \rho_{\text{exe}} < \rho_{\text{sem}} < 0$ to penalize failures according to their severity. Specifically, $\texttt{ExtractionFail}(q')$ indicates cases where the model fails to generate a recognizable SQL block (i.e., $q'$ is empty or null), while $\texttt{ExecError}(q')$ captures valid SQL strings that fail execution (e.g., syntax errors). For valid rewrites, the performance gain is quantified by an asymmetric scaling function $\mathcal{F}$:
\begin{equation}
    \mathcal{F}(T_{q'}, T_0) = 
    \begin{cases}
        \eta \cdot \tanh(\log(T_0 / T_{q'})), & \text{if } T_{q'} < T_0, \\
        \tanh(\log(T_0 / T_{q'})), & \text{otherwise}.
    \end{cases}
\end{equation}
Here, $\eta > 1$ acts as an incentive factor to amplify gradients for positive speedups, encouraging aggressive optimization over conservative equivalence.

\begin{figure}[t]
    \centering
    \includegraphics[width=0.998\linewidth]{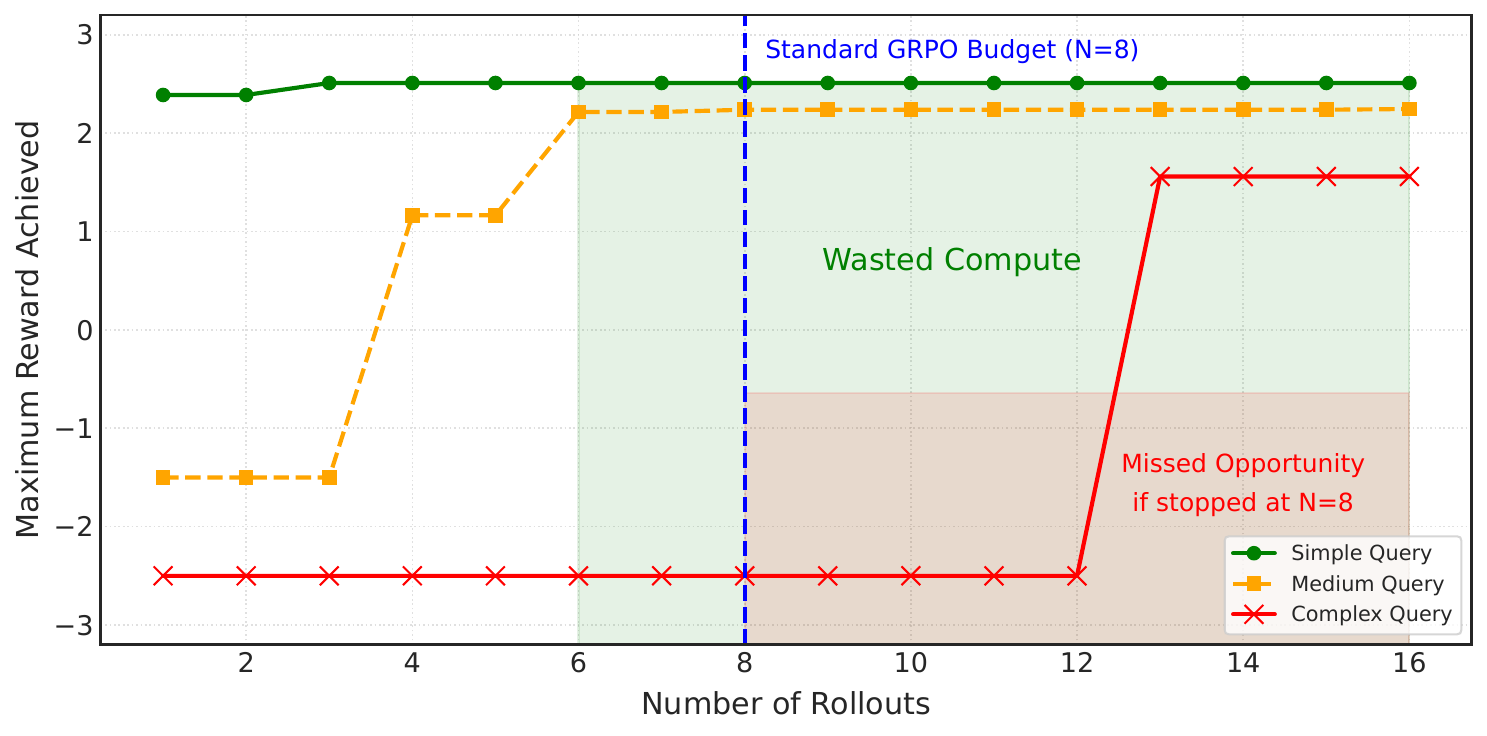} 
    \vspace{-3.5mm}
    \caption{Impact of Rollout Budget on Optimization Discovery. Simple queries (Green) saturate early, causing compute waste, whereas complex queries (Red) require extended exploration budgets to discover valid optimization paths.} 
    \label{fig:budget_impact}
    \vspace{-2mm}
\end{figure}

\noindent
\textbf{Complexity-Adaptive Dynamic Rollout.}
Standard GRPO allocates a fixed rollout budget $N$ to every prompt. However, this uniform approach lacks flexibility given the significant heterogeneity in SQL query complexity. As illustrated in Figure~\ref{fig:budget_impact}, simple queries tend to saturate rewards early, resulting in significant computational waste. Conversely, highly complex queries often fail to produce a single valid rewrite under a limited budget, requiring extended exploration to satisfy validity constraints and discover effective optimization paths.

To optimize resource allocation, we propose a two-stage budgeting mechanism termed Complexity-Adaptive Dynamic Rollout. In the initial pilot phase, we assign a minimal budget $k_\text{pilot}$ to all prompts to gather performance statistics. Based on these metrics, we determine the allocation weight $W_i$ for each prompt $i$ via:
\begin{equation}
    W_i = \alpha \cdot \mathbb{I}[\max(R_{\text{pilot}}) < \rho_{\text{sem}}] + \beta \cdot \tilde{\mathbb{H}}(\pi) + \gamma \cdot \tilde{\sigma}^2_{R}.
\end{equation}

Here, the indicator function $\mathbb{I}[\cdot]$ acts as a failure detector, explicitly prioritizing instances that fail to yield valid SQL rewrites during the pilot phase. To quantify exploration value, we incorporate normalized policy entropy $\tilde{\mathbb{H}}(\pi)$ and reward variance $\tilde{\sigma}^2_{R}$ as proxies for model uncertainty and optimization potential. 
Specifically, the policy entropy $\mathbb{H}(\pi)$ measures the token-level average entropy, capturing the model's confidence in its rewrite trajectory~\cite{arpo, seed-grpo}. High entropy suggests the model is oscillating between multiple rewrite paths, indicating a complex landscape that warrants further exploration. Similarly, the reward variance $\tilde{\sigma}_{R}^{2}$ captures the sensitivity of the query to minor structural changes. Instances with high variance indicates a rugged solution space, justifying a larger rollout budget to stabilize gradient estimation.
Finally, the remaining computational budget is allocated proportionally to $W_i$, directing resources toward long-tail instances that demand intensive search.

\begin{figure}[!t]
    \centering
    \includegraphics[width=0.98\linewidth]{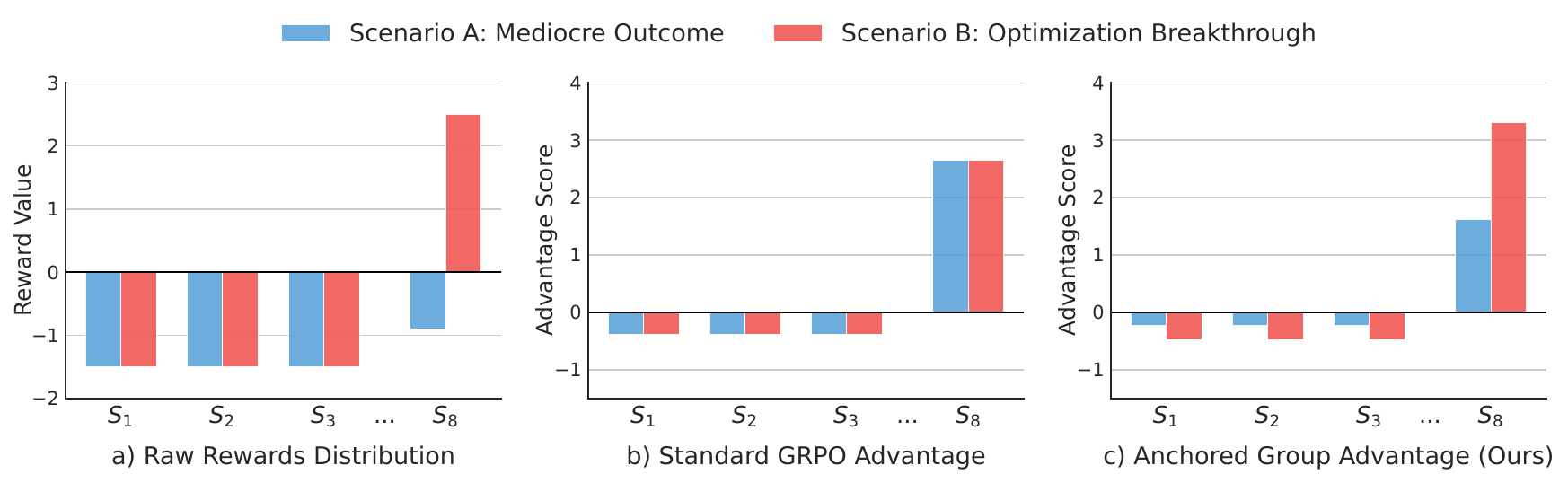} 
    \vspace{-2mm}
    \caption{Comparison of Advantage Estimation. Standard normalization (b) creates a false equivalence between the mediocre survivor in Scenario A and the genuine breakthrough in Scenario B. Our Anchored Group Advantage (c) uses absolute anchors to suppress this inflated signal while correctly amplifying the optimization success.}
    \label{fig:advantage_comparison}
    \vspace{-4mm}
\end{figure}

\noindent
\textbf{Anchored Group Advantage.}
Standard GRPO computes advantages using intra-group Z-score normalization. However, in the context of SQL optimization, this strict relative scoring ignores the absolute quality of the rewrites. This limitation often leads to false equivalence in advantage assignment. As illustrated in Figure~\ref{fig:advantage_comparison}, standard normalization induces a misleading parity. Specifically, it assigns a similarly high advantage score to a mediocre survivor in a low-quality group (Scenario A), comparable to that of a genuine optimization breakthrough in a high-performing group (Scenario B). Consequently, within a group dominated by syntax errors, a valid yet slow rewrite receives a high positive advantage merely for being executable. This behavior erroneously reinforces performance deterioration, as the model is rewarded for producing suboptimal SQL simply because it outperforms catastrophic failures rather than achieving genuine optimization.
To rectify this, we propose Anchored Group Advantage, which fuses relative rank with absolute performance anchors:
\begin{equation}
\begin{aligned}
    A_\text{anchor}^{(i)} &= (1-\lambda) \underbrace{\frac{r_i - \mu_G}{\sigma_G + \epsilon}}_{\text{Relative}} + \lambda \underbrace{\frac{r_i - b}{S} \sqrt{G}}_{\text{Absolute}}, \\
    \hat{A}^{(i)} &= A_{\text{anchor}}^{(i)} - \frac{1}{G} \sum_{j=1}^{G} A_\text{anchor}^{(j)}.
    \label{eq:anchor}
\end{aligned}
\end{equation}
The absolute term incorporates a global baseline $b$ and a reward scaling factor $S$, explicitly weighted by $\sqrt{G}$ to match the magnitude of the Z-score normalized relative term. This mechanism ensures that when a group performs poorly overall, the negative absolute component dominates to suppress misleading relative signal. The final centering step maintains the zero-sum property required for stable policy gradient updates.

\noindent
\textbf{Optimization Objective.}
Our training objective integrates the Anchored Group Advantage into the GRPO framework, incorporating a KL divergence penalty to constrain the policy within the linguistic distribution of the reference model $\pi_\text{ref}$ (i.e., the SFT model). The final loss function is defined as:
\begin{equation}
    \mathcal{L}(\theta) = -\mathbb{E}_{q \sim P(Q), o \sim \pi_{\theta}(q)} \left[ r_t(\theta) \hat{A} - \sigma D_{KL}(\pi_{\theta} || \pi_\text{ref}) \right],
\end{equation}
where $r_t(\theta) = \frac{\pi_{\theta}(o|q)}{\pi_{\theta_{\text{old}}}(o|q)}$ denotes the probability ratio between the current policy $\pi_{\theta}$ and the old policy $\pi_{\theta_{\text{old}}}$. The term $\hat{A}$ represents the Anchored Group Advantage calculated in Eq.~\ref{eq:anchor}. The hyperparameter $\sigma$ serves as the coefficient for the KL divergence penalty, balancing reward optimization with generation diversity.

\subsection{Verification-Driven Self-Correction}
To guarantee the practical executability of generated rewrites, we implement a verification-driven self-correction mechanism that integrates database diagnostics directly into the inference loop.

Upon generating a candidate rewrite, we first subject it to the DBMS's \texttt{EXPLAIN} command. This utility acts as a zero-cost syntax filter, validating the structural integrity of the query without triggering the high latency associated with actual data retrieval. If the candidate passes this check, it is finalized immediately.

In cases where \texttt{EXPLAIN} returns an error, we trigger a regenerative repair process. We construct an augmented context containing the original slow query, the invalid candidate, and the specific error message returned by the DBMS. This diagnostic feedback allows the model to pinpoint syntactic faults (e.g., keyword misuse or column ambiguity) and perform self-correction. This closed-loop approach significantly enhances robustness, ensuring that the system recovers from parsing failures autonomously.

\section{Experiments}
\label{sec:experiments}

In this section, we conduct extensive experiments to demonstrate the efficiency of our rewrite models from multiple perspectives, including detailed analysis of query execution latency on several benchmarks, rewrite cost, data-aware analysis and ablation study.

\begin{table}[!t]
    \centering
    \caption{Overall statistics across different benchmarks.}
    \vspace{-3mm}
    \resizebox{0.48\textwidth}{!}{
        \begin{tabular}{lcccc}
        \toprule
        Benchmark & \# Templates & \# Tokens & \# Predicates & \# SubQueries \\
        \midrule
        TPC-DS~\cite{tpcds} & 103 & 390.18 & 15.94 & 1.63 \\
        TPC-H~\cite{tpch} & 22 & 148.23 & 7.55 & 0.68  \\
        DSB~\cite{dsb} & 37 & 433.11 & 21.19 & 1.35 \\
        Calcite~\cite{calcite} & 793 & 87.38 & 2.42 & 1.56 \\
        SQL-Factory~\cite{sql-factory} & 50k & 76.18 & 2.23 & 0.04 \\
        \midrule
        SQL-MCTS & 11.7k & 334.47 & 11.24 & 1.89 \\
        \bottomrule
        \end{tabular}
    }
    \vspace{-1.5mm}
\label{tab:dataset}
\end{table}

\subsection{Experiment Setup}

\subsubsection{Environment}
The experiments are conducted on a PostgreSQL 13 server hosted on the VolcEngine platform~\cite{volcengine}, featuring a configuration with 2 cores and 4GB of RAM. For model training, we utilize a GPU server equipped with 8 NVIDIA A800-SXM4-80GB GPUs and an Intel(R) Xeon(R) Platinum 8336C CPU.

\subsubsection{Benchmarks}
We evaluate the performance of our query rewrite model using four typical database benchmarks.
(1)
\textbf{TPC-DS}~\cite{tpcds} is an industry-standard benchmark designed for decision support systems, consisting of 24 tables and 103 queries.
(2)
\textbf{TPC-H}~\cite{tpch} is also a well-known benchmark with 8 tables and 22 query templates.
(3)
\textbf{DSB}~\cite{dsb} is adapted from TPC-DS,featuring a complex data distribution and challenging query templates, with a total of 37 query templates.
(4)
\textbf{Calcite}~\cite{calcite} is a real-world benchmark mainly used to evaluate rewrite rules, comprising 6 tables and a total of 793 queries, from which we randomly select 50 queries.
We used the official tools of these four benchmarks to generate 10GB of data for experimentation.

\subsubsection{Baseline}
\begin{sloppypar}
We compare our trained query rewrite model with both rule-based and LLM-based approaches.
\end{sloppypar}

\noindent
\textbf{Rule-based Methods.} We compare three existing rule-based methods. (1) LearnedRewrite~\cite{learnedrewrite} utilizes a MCTS algorithm coupled with a learned cost estimation model to explore the space of rewrite rule orders. (2) LLM-R$^2$~\cite{llm-r2} employs LLM with In-Context Learning ability to select rewrite rules. (3) R-Bot~\cite{r-bot} implements a retrieval-augmented generation (RAG) approach to select rewrite rules.

\noindent
\textbf{LLM-based methods.} We also compare our model against naive LLM, such as DeepSeek-R1~\cite{deepseek-r1} and GPT-4o~\cite{gpt-4o}. These models directly generate query rewrites without the use of rewrite rules. Additionally, we include a baseline trained using a naive GRPO method to represent the E$^3$-Rewrite~\cite{e3} approach. We chose not to compare with QUITE~\cite{quite}, as its complex agent-based architecture and lack of open-source availability prevent a fair comparison.

\subsubsection{Evaluation Metrics} 
(1) Query Latency. The time taken to execute a query, measured as average, median, 75th, and 95th percentile latencies. To ensure stability, each query undergoes one warm-up run followed by three executions and recording the mean latency. Besides, queries exceeding 300 seconds are considered timeouts.
(2) Equivalence Rate. The proportion of rewritten queries that are semantically equivalent to the original queries, verified by executing both queries and comparing whether their results are identical.

\begin{table}[!t]
    \centering
    \caption{Human evaluation across different benchmarks.}
    \vspace{-3mm}
    \resizebox{0.48\textwidth}{!}{
        \begin{tabular}{lccc}
        \toprule
        Benchmark & Opt. Non-Triviality & Struct. Complexity & Sem. Coherence \\
        \midrule
        TPC-DS~\cite{tpcds} & 4.25 & 3.42 & 4.96 \\
        TPC-H~\cite{tpch} & 3.54 & 2.77 & 5.00  \\
        DSB~\cite{dsb} & 4.29 & 3.78 & 5.00 \\
        Calcite~\cite{calcite} & 3.10 & 1.81 & 3.37 \\
        SQL-Factory~\cite{sql-factory} & 3.46 & 2.07 & 4.70 \\
        \midrule
        SQL-MCTS (All) & 4.26 & 2.93 & 4.92 \\
        SQL-MCTS (Top 70\%) & 4.74 & 3.36 & 4.97 \\
        \bottomrule
        \end{tabular}
    }
    \vspace{-1.5mm}
\label{tab:human_eval}
\end{table}

\subsubsection{Implementation Details}
We use DeepSeek-V3~\cite{deepseek-v3} to generate slow queries within the Monte Carlo Tree Search. For the base model, we utilize the Qwen3~\cite{qwen3} series, including Qwen3-8B and Qwen3-14B. The models are trained using the Verl~\cite{verl} framework, with both SFT and GRPO learning rates set to $3e^{-6}$, a batch size of 256, and 1 epoch. Other configurations are maintained according to the Verl setup. For the GRPO training phase, we set the temperature to 1.0 to encourage diverse exploration. For all other inference tasks, we set the temperature to 0.0 to ensure output stability. 
Additionally, in the reward function of SQL-GRPO, we assign additional negative rewards $\rho_{\text{fmt}}=-3$, $\rho_{\text{exe}}=-2.5$, and $\rho_{\text{sem}}=-1.5$. For rewritten queries that achieve positive latency improvement while preserving equivalence, we provide a high bonus $\eta=3$. Accordingly, the scaling factor $S$ in Eq.~\ref{eq:anchor} is set to 3 to align with the reward range. Crucially, we set the global baseline $b=0$ to enforce absolute correctness constraints. All remaining aggregation coefficients, including $\alpha$, $\beta$, $\gamma$, and $\lambda$, are set to equal weights. The entire training process, including both SFT and GRPO phases, was completed in approximately 34.9 hours.

\begin{table*}[!th]
    \centering
    \caption{Comparison of Query Latency across TPC-DS schema-based benchmarks.}
    \vspace{-3mm}
    \resizebox{\textwidth}{!}{
    \begin{tabular}{lccccccccccccccc}
    \toprule
    \multirow{2}{*}{Method} & \multicolumn{5}{c}{SQL-MCTS (10G)} & \multicolumn{5}{c}{TPC-DS (10G)} & \multicolumn{5}{c}{DSB (10G)} \\
    \cmidrule(lr){2-6} \cmidrule(lr){7-11} \cmidrule{12-16}
    & Mean & Median & 75th & 95th & Equi. & Mean & Median & 75th & 95th & Equi. & Mean & Median & 75th & 95th & Equi. \\
    \midrule
    Original             & 69.73 & 29.19 & 69.14 & 300.00 & - & 49.00 & 10.60 & 36.57 & 300.00 & - & 44.88 & 11.42 & 20.75 & 300.00 & - \\
    \midrule
    LearnedRewrite       & 67.03 & 29.89 & 68.69 & 297.16 & 70\%  & 46.72 & 11.49 & 35.69 & 300.00 & 71\%  & 38.22 & 10.92 & 20.72 & 300.00 & 43\%  \\
    LLM-R$^2$ (DeepSeek-R1) & 67.96 & 27.08 & 69.20 & 300.00 & 50\%  & 46.05 & 11.05 & 33.10 & 300.00 & 66\%  & 14.33 & 10.20 & 16.54 & 50.47  & 83\%  \\
    LLM-R$^2$ (GPT-4o)      & 68.08 & 27.11 & 66.43 & 300.00 & 50\%  & 49.11 & 10.71 & 41.35 & 300.00 & 69\%  & 29.21 & 10.03 & 17.57 & 126.20 & 70\%  \\
    R-Bot (DeepSeek-R1)  & 58.79 & 22.05 & 68.48 & 297.16 & 46\%  & 44.66 & 10.87 & 37.92 & 297.35 & 69\%  & 21.45 & 12.44 & 21.18 & 81.76  & \textbf{94\%}  \\
    R-Bot (GPT-4o)       & 51.09 & 23.61 & 53.79 & 191.32 & 38\%  & 42.40 & 10.34 & 33.92 & 293.49 & 69\%  & 20.37 & 10.42 & 21.08 & 69.64  & \underline{91\%}  \\
    \midrule
    DeepSeek-R1          & \underline{13.40} & 9.05  & 12.63 & \textbf{31.17}  & \underline{86\%}  & \textbf{26.87} & 9.57  & \textbf{20.96} & \underline{106.58} & \underline{72\%}  & 19.03 & \underline{8.80}  & 16.42 & 45.21  & 86\%  \\
    GPT-4o               & 25.36 & 8.98  & 13.43 & 107.53 & 85\%  & 33.31 & 9.91  & 26.02 & 225.21 & 51\%  & 19.72 & 9.57  & \underline{15.78} & 67.37  & 48\%  \\
    Qwen3-8B             & 38.72 & 9.13  & 26.80 & 270.20 & 65\%  & 34.92 & 9.67  & 27.35 & 255.07 & 46\%  & 35.58 & 9.52  & 17.75 & 300.00 & 37\%  \\
    Qwen3-14B            & 38.13 & 9.02  & 24.40 & 263.69 & 65\%  & 35.77 & 9.56  & 24.88 & 293.49 & 46\%  & 19.75 & 9.28  & 17.02 & 67.37  & 62\%  \\
    \midrule
    LASER-8B               & 15.12 & \textbf{7.95}  & \underline{12.30} & 34.31  & 85\%  & 28.03 & \underline{9.55}  & 24.38 & 112.52 & 64\%  & \underline{10.56} & 9.04  & 16.91 & \underline{37.30}  & 72\%  \\
    LASER-14B              & \textbf{13.03} & \underline{8.98}  & \textbf{12.09} & \underline{31.63}  & \textbf{90\%}  & \underline{27.16} & \textbf{9.53}  & \underline{21.06} & \textbf{104.24}  & \textbf{77\%}  & \textbf{10.33} & \textbf{7.84}  & \textbf{15.78} & \textbf{25.88}  & 83\% \\
    \bottomrule
    \end{tabular}
    }
    \vspace{-1.5mm}
\label{tab:methods}
\end{table*}

\subsection{Quality Evaluation of SQL-MCTS}

To demonstrate the superior quality of SQL-MCTS, we compare it against established benchmarks and the query generation framework SQL-Factory~\cite{sql-factory}. As presented in Table~\ref{tab:dataset}, our analysis focuses on key complexity metrics including the number of templates, token count, predicates, and subqueries.
The analysis reveals that SQL-MCTS strikes a distinct balance between query diversity and structural complexity. While SQL-Factory generates a vast number of templates, they are notably simplistic with only 76.18 tokens and 0.04 subqueries. In contrast, SQL-MCTS maintains high complexity across 11.7k templates. Although TPC-DS exhibits slightly higher token counts, it is limited to a narrow set of 103 templates. By combining structural depth with extensive scale, SQL-MCTS provides a challenging environment for training SQL rewrite models.

To further strictly validate that these statistical metrics translate into genuine optimization challenges, we conducted a blind human evaluation. We randomly sampled 100 queries from the SQL-MCTS and SQL-Factory and compared them against all benchmarks. 
Three PhD-level STEM students with SQL experience evaluated these anonymized queries on a 5-point scale across three dimensions. Specifically, we defined \textit{Optimization Non-Triviality} as the extent to which a query necessitates high-level architectural refactoring rather than trivial syntactic cleanup. \textit{Structural Complexity} isolates logical depth, such as rigid join topologies and deep nesting, from mere physical resource consumption. Finally, \textit{Semantic Coherence} measures the plausibility of the business logic to ensure the generated SQL is not random syntactic noise.
First, as shown in Table~\ref{tab:human_eval}, the full SQL-MCTS dataset (\textit{All}) already demonstrates high quality, achieving an Optimization Non-Triviality score of 4.26. This performance is statistically on par with the industry-standard TPC-DS (4.25) and significantly outperforms other baselines.
To further ensure a rigorous comparison against the most challenging workloads, we isolated the \textit{Refined Subset} (Top 70\% by complexity), filtering out structurally simpler instances reserved for distributional regularization during training.
The evaluation results confirm that this subset presents a significantly harder optimization landscape. Specifically, the Refined Subset achieves an Optimization Non-Triviality score of 4.74, surpassing all benchmarks.
Crucially, it maintains a Semantic Coherence score of 4.97, which is indistinguishable from expert-crafted benchmarks, ensuring that the increased complexity does not compromise logical validity.

\subsection{Performance Comparison}

\noindent
\textbf{Query Execution Latency.} 
The first comparison focuses on query latency across TPC-DS, DSB, and our generated SQL-MCTS benchmarks, all utilizing the same schema. As shown in Table~\ref{tab:methods}, the performance of our LASER model shows significant improvements after training, achieving remarkable reductions in query latency while maintaining high query equivalence.

Specifically, the complexity of SQL-MCTS results in difficulties for all rule-based methods, as their limited set of rules cannot effectively handle such complex queries. This limitation leads to low query efficiency and equivalence for these methods. Similarly, the original Qwen3 model struggles to manage these complex queries. However, after training with our approach, the performance of the LASER model has drastically improved. The 8B model now performs at the level of larger models like DeepSeek-R1 and GPT-4o, while the 14B model even surpasses DeepSeek-R1. For example, the average execution time for LASER-14B was optimized from 69.73 seconds to 13.03 seconds, achieving 90\% equivalence.
Besides, our model reduces the average query time from 49.00 seconds to 27.16 seconds on TPC-DS and from 44.88 seconds to 10.33 seconds on DSB. These improvements are substantial, and our method achieves a state-of-the-art in terms of performance, far surpassing the existing rule-based methods. Furthermore, the performance of our LASER model exceeds the capabilities of API-based solutions, which typically struggle to provide similar reductions in execution time while maintaining high equivalence.

This performance improvement demonstrates that with proper training using high quality training datasets, even the 8B and 14B models can outperform larger, more complex systems, making them highly efficient for complex query rewrite tasks.

\begin{table}[!t]
    \centering
    \caption{Cost analysis on DSB benchmark (10GB scale).}
    \vspace{-3mm}
    \resizebox{0.48\textwidth}{!}{
    \begin{tabular}{lcccc}
    \toprule
    Method               & Times (s) & Cost (\$) & Memory (GB)        & Localizable \\
    \midrule
    LLM-R$^2$ (DeepSeek-R1) & 100.5     & 23.7      & >1000 & $\times$           \\
    LLM-R$^2$ (GPT-4o)      & 43.9      & 60.0      & -                  & $\times$           \\
    R-Bot (DeepSeek-R1)  & 982.3     & 243.9     & >1000 & $\times$           \\
    R-Bot (GPT-4o)       & 169.8     & 2075.5    & -                  & $\times$           \\
    DeepSeek-R1          & 130.2     & 31.2      & >1000 & $\times$          \\
    GPT-4o               & 18.5      & 167.7     & -                  & $\times$           \\
    \midrule
    LASER-8B          & 24.6      & 5.92      & 32                 & \checkmark           \\
    LASER-14B         & 28.8      & 15.6      & 40                 & \checkmark           \\
    \bottomrule
    \end{tabular}
    }
    \vspace{-1.5mm}
\label{tab:cost}
\end{table}

\noindent
\textbf{Cost Analysis.}
To provide a fair comparison, we focus on the DSB Benchmark to further analyze the inference costs associated with these methods. As shown in Table~\ref{tab:cost}, Time represents the average query rewrite time, Cost includes both the API call costs and GPU server rental costs~\cite{volcengine, autodl}, scaled to 10,000 queries for better clarity. Memory indicates the minimum required VRAM for inference, and Localizable denotes the feasibility of performing inference locally.

The analysis demonstrates that despite achieving high performance, LASER models exhibit significantly faster rewrite times. This is especially notable compared to R-Bot, which requires a stepwise recommendation of rule applications. Moreover, the compact scale of LASER enables efficient deployment on local machines with minimal VRAM requirements. This efficiency not only reduces inference time but also minimizes GPU operational costs. Additionally, the ability to run our models locally without relying on external APIs means that there are no hidden costs associated with data uploads, and privacy concerns are mitigated.

\begin{table*}[!t]
    \centering
    \caption{Comparison of Query Latency on Unseen Schema Benchmarks (TPC-H and Calcite).}
    \vspace{-3mm}
    \small
    \begin{tabular*}{\textwidth}{@{\extracolsep{\fill}\hspace{\tabcolsep}}lcccccccccc@{\hspace{\tabcolsep}}}
    \toprule
    \multirow{2}{*}{Method} & \multicolumn{5}{c}{TPC-H (10G)} & \multicolumn{5}{c}{Calcite (10G)} \\
    \cmidrule(lr){2-6} \cmidrule(lr){7-11}
    & Mean & Median & 75th & 95th & Equi. & Mean & Median & 75th & 95th & Equi. \\
    \midrule
    Original             & 77.43          & 30.59          & 50.59          & 300.00         & -              & 32.59         & 7.06          & 16.01          & 300.00         & -             \\
    \midrule
    LearnedRewrite       & 65.98          & 31.47          & 41.57          & 300.00         & 68\%           & 32.67         & 8.45          & 18.94          & 300.00         & 75\%          \\
    LLM-R$^2$ (DeepSeek-R1) & 68.19          & 30.67          & 52.22          & 300.00         & 95\%           & 33.79         & 8.35          & 19.68          & 300.00         & \underline{81\%}    \\
    LLM-R$^2$ (GPT-4o)      & 64.11          & 27.41          & 40.27          & 300.00         & 90\%           & 26.97         & 6.92          & 14.97          & 113.96         & 72\%          \\
    R-Bot (DeepSeek-R1)  & 41.90          & 30.22          & 38.90          & 74.64          & 90\%           & 23.71         & 7.49          & 17.23          & 76.80          & 79\%          \\
    R-Bot (GPT-4o)       & 42.25          & 30.18          & 39.59          & 74.23          & \underline{95\%}     & 17.68         & 5.29          & 14.62          & 75.32          & 81\% \\
    \midrule
    DeepSeek-R1          & 38.70          & \underline{27.73}    & \textbf{32.83} & \textbf{52.65} & 86\%           & 13.00         & 6.92          & 12.08          & \underline{24.88}    & \textbf{89\%}          \\
    GPT-4o               & 41.17          & 29.65          & 40.88          & 77.29          & 90\%           & 14.21         & \textbf{5.50} & 12.43          & 50.22          & 51\%          \\
    Qwen3-8B             & 52.13          & 29.68          & 57.77          & 288.74         & 90\%           & 20.63         & 6.53          & 12.42          & 76.33          & 62\%          \\
    Qwen3-14B            & 40.45          & 27.90          & 39.78          & 73.18          & 90\%           & 17.60         & 7.41          & 12.59          & 75.90          & 62\%          \\
    \midrule
    LASER-8B               & \underline{34.61}    & 28.63          & 44.98          & 71.36          & 90\%           & \underline{11.90}   & 6.62          & \underline{11.87}    & 39.95          & 65\%          \\
    LASER-14B              & \textbf{31.78} & \textbf{26.73} & \underline{38.14}    & \underline{70.50}    & \textbf{100\%} & \textbf{9.98} & \underline{5.68}    & \textbf{11.33} & \textbf{24.11} & 75\% \\
    \bottomrule
    \end{tabular*}
    \vspace{-1.5mm}
\label{tab:cross}
\end{table*}

\subsection{Transferability Across Benchmarks}

Due to our SQL-MCTS dataset being generated using the TPC-DS schema, we also evaluate the transferability of our LASER models on other benchmarks, including TPC-H and Calcite, which were not part of the training process. This allows us to assess how well our model generalizes to new, unseen database schemas.

As shown in Table~\ref{tab:cross}, our LASER-8B and LASER-14B models demonstrate excellent performance on both the TPC-H and Calcite benchmarks. In particular, the LASER-14B model achieves the best results across all metrics, with a mean query execution time of 31.78 seconds on TPC-H and 9.98 seconds on Calcite, both of which represent significant improvements compared to traditional rule-based methods. 
Additionally, LASER-8B also performs well, making it a competitive option even compared to larger models.
This demonstrates the transferability of our approach, where our model not only performs well on the benchmark with the schema it was trained on but also generalizes effectively to other schema benchmarks. The LASER models are thus highly adaptable and capable of achieving state-of-the-art performance across a variety of database schemas, without needing extensive retraining.

\begin{table*}[!t]
    \centering
    \caption{Data-Aware Analysis on DSB Benchmark.}
    \vspace{-3mm}
    \small
    \begin{tabular*}{\textwidth}{@{\extracolsep{\fill}\hspace{\tabcolsep}}lcccccccccccc@{\hspace{\tabcolsep}}}
    \toprule
    \multirow{2}{*}{Method} & \multicolumn{4}{c}{DSB (1G)} & \multicolumn{4}{c}{DSB (10G)} & \multicolumn{4}{c}{DSB (50G)} \\
    \cmidrule(lr){2-5} \cmidrule(lr){6-9} \cmidrule(lr){10-13}
    & Mean & Median & 75th & 95th & Mean & Median & 75th & 95th & Mean & Median & 75th & 95th \\
    \midrule
    Original & 8.64 & 0.27 & 0.69 & 24.67 & 44.88 & 11.42 & 20.75 & 300.00    & 82.05 & 15.52 & 111.14 & 300.00 \\
    \midrule
    DeepSeek-R1 & 0.62 & 0.22 & 0.54 & 1.91  & 19.03 & \underline{8.80}  & 16.42 & 45.21  & \underline{46.34} & 20.48 & 56.44  & 300.00 \\
    GPT-4o & 0.41 & \underline{0.17} & \underline{0.36} & 1.46  & 19.72 & 9.57  & \underline{15.78} & 67.37  & 63.72 & 12.32 & 40.02  & 300.00 \\
    Qwen3-8B & 5.96 & 0.24 & 0.63 & 3.10  & 35.58 & 9.52  & 17.75 & 300.00 & 71.11 & 14.76 & 46.06  & 300.00 \\
    Qwen3-14B & 1.37 & 0.23 & 0.58 & 1.23  & 19.75 & 9.28  & 17.02 & 67.37  & 57.54 & 13.19 & 41.00  & 300.00 \\
    \midrule
    LASER-8B & \underline{0.34} & 0.21 & 0.46 & \underline{1.02}  & \underline{10.56} & 9.04  & 16.91 & \underline{37.30}  & 50.05 & \underline{11.89} & \underline{39.26}  & 300.00 \\
    LASER-14B & \textbf{0.33} & \textbf{0.16} & \textbf{0.34} & \textbf{0.95}  & \textbf{10.33} & \textbf{7.84}  & \textbf{15.78} & \textbf{25.88}  & \textbf{39.63} & \textbf{10.76} & \textbf{33.55}  & \textbf{185.29} \\
    \bottomrule
    \end{tabular*}
    \vspace{-2mm}
\label{tab:scale}
\end{table*}

\subsection{Data-aware Analysis}
In this experiment, we further evaluate the performance of our LASER models on the DSB dataset at three different scales: 1G, 10G, and 50G, which represent small, medium, and large database sizes. The results, shown in Table~\ref{tab:scale}, demonstrate that our models maintain high performance even as the dataset size increases.
For example, on DSB (50G), the LASER-14B model achieves a mean execution time of 39.6 seconds, significantly faster than other methods. 
These results show that our LASER models are highly effective in handling increasing data volumes, making them suitable for large-scale database environments.

\subsection{Ablation Study}
In this ablation study, we evaluate the impact of various components of our LASER approach by progressively removing key elements. The results, as shown in Table~\ref{tab:ablation}, highlight the importance of each component in achieving optimal performance.

First, without Supervised Fine-Tuning, the model suffers from a lack of a structured initialization, leading to poorer training stability and a lower training ceiling. Specifically, the absence of SFT causes the Equivalence Rate to plummet from 83\% to 64\%, while the mean execution time concurrently increases to 14.24 seconds.
Next, removing the Complexity-Aware Dynamic Rollout leads to inefficiencies in the training process. Without this feature, the model tends to waste training iterations on less important samples while insufficiently exploring challenging ones. This significantly hampers performance on complex queries, as evidenced by the 95th percentile latency rising to 36.86 seconds.
The absence of Anchored Group Advantage further exacerbates the issue. Without this component, the model may assign disproportionately high advantages to poorly performing samples, negatively impacting the learning process, further demonstrating the importance of this mechanism.
When comparing our method to a naive GRPO (E$^3$-Rewrite) setup, which uses the most basic GRPO strategy, the performance significantly deteriorates.
Most notably, the 95th percentile latency spikes to nearly 2.5 times that of LASER-14B. This sharp disparity indicates that standard GRPO struggles to navigate the sparse solution space of structurally complex queries. In contrast, our approach effectively tackles these hard samples, ensuring robust optimization even for the most challenging inputs.
Finally, eliminating the Verification-Driven Self-Correction mechanism compromises the system's robustness. Specifically, the Equivalence Rate declines to 78\%, and the mean latency increases to 11.78 seconds. 

In summary, the ablation study underscores the critical role of each component in the LASER framework. Removing any of these elements leads to a visible decline in performance, both in terms of query execution latency and query equivalence.

\begin{table}[!t]
    \centering
    \caption{Ablation Study on DSB Benchmark.}
    \vspace{-3mm}
    \resizebox{0.48\textwidth}{!}{
        \begin{tabular}{lccccc}
        \toprule
        Method & Mean & Median & 75th & 95th & Equi. \\
        \midrule
        Original & 44.88 & 11.42 & 20.75 & 300.00 & - \\
        LASER-14B & \textbf{10.40} & \textbf{7.84} & \textbf{15.78} & \textbf{25.88} & \textbf{83\%} \\
        \midrule
        w/o SFT & 14.24 & 10.24 & 17.75 & 45.33 & 64\% \\
        w/o Dynamic Rollout & 16.57 & 9.30 & 16.42 & 36.86 & 70\% \\
        w/o Anchored Group Adv. & 13.40 & 9.01 & 18.42 & 31.97 & 75\% \\
        Naive GRPO & 18.57 & 10.03 & 16.81 & 53.63 & 67\% \\
        w/o Self-Correction & 11.78 & 8.34 & 17.78 & 28.08 & 78\% \\
        \bottomrule
        \end{tabular}
    }
    \vspace{-2mm}
\label{tab:ablation}
\end{table}

\subsection{Cross-Database Evaluation on MySQL}
We also tested the performance of our LASER models on MySQL using the DSB Benchmark, as all prior experiments were conducted on PostgreSQL. The results, shown in Table~\ref{tab:mysql}, indicate that our models perform equally well on MySQL, achieving results comparable to DeepSeek-R1. This experiment serves as a crucial validation against a different native query optimizer, confirming that LASER's optimization capabilities are not overfitted to PostgreSQL's planner logic but represent generalized SQL optimization rules.
Specifically, the LASER-8B and 14B models show significant improvements over the original queries and exhibit competitive performance against the DeepSeek-R1 model. For example, LASER-14B achieves a mean query execution time of 30.49 seconds, while DeepSeek-R1 performs at 28.71 seconds. The equivalence for LASER-14B is also higher than DeepSeek-R1, demonstrating the robustness and flexibility of our approach across different database systems.

These results confirm that LASER is not only effective on PostgreSQL but also generalizes well to other database platforms, making it a versatile solution for various database environments.

\begin{table}[!t]
    \centering
    \caption{Cross-database evaluation using MySQL on the DSB benchmark (10GB scale).}
    \vspace{-3mm}
    \small
    \begin{tabular*}{\linewidth}{@{\extracolsep{\fill}\hspace{\tabcolsep}}lccccc@{\hspace{\tabcolsep}}}
    \toprule
    Method       & Mean           & Median         & 75th           & 95th            & Equi.         \\
    \midrule
    Original     & 64.87          & 18.29          & 69.18          & 300.00          & -             \\
    \midrule
    DeepSeek-R1  & \textbf{28.71} & \textbf{11.07} & \textbf{35.58} & \underline{118.10}    & \underline{75\%}    \\
    DeepSeek-V3  & 48.23          & 16.42          & 47.14          & 184.87          & 70\%          \\
    GPT-4o       & 37.98          & 15.02          & 42.50          & 125.92          & 64\%          \\
    Qwen3-8B     & 56.93          & 18.29          & 67.76          & 300.00          & 45\%          \\
    Qwen3-14B    & 38.50          & 12.25          & 45.55          & 144.01          & 56\%          \\
    \midrule
    LASER-8B  & 37.69          & 11.84          & 44.16          & 133.43          & 70\%          \\
    LASER-14B & \underline{30.49}    & \underline{11.56}    & \underline{42.24}    & \textbf{105.09} & \textbf{78\%} \\
    \bottomrule
    \end{tabular*}
    \vspace{-2mm}
\label{tab:mysql}
\end{table}

\begin{figure}[!t]
    \centering
    \includegraphics[width=0.996\linewidth]{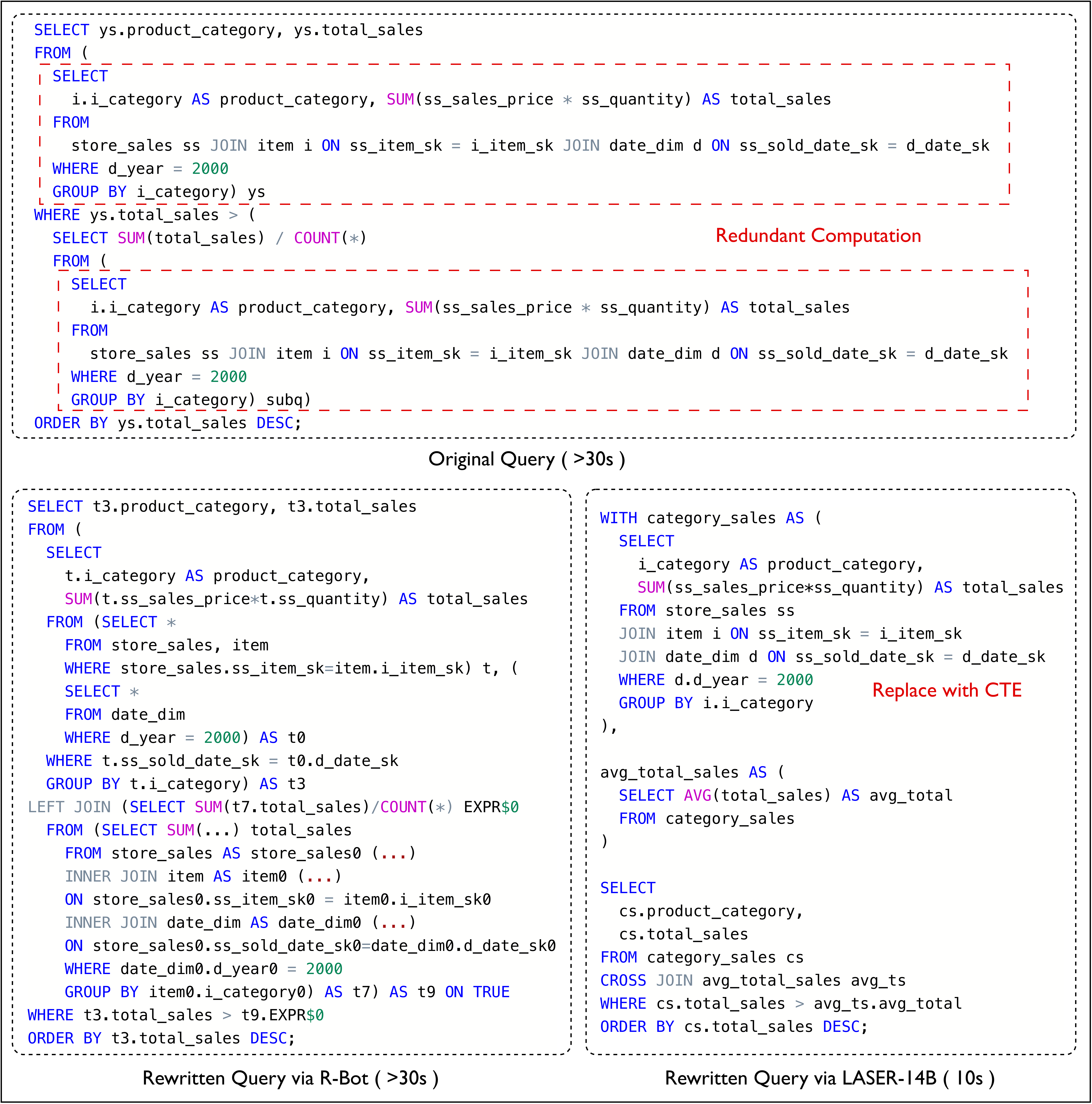} 
    \vspace{-3.5mm}
    \caption{An Example of generated slow queries.} 
    \label{fig:case}
    \vspace{-4mm}
\end{figure}

\subsection{Case Study}

To demonstrate how LASER optimizes query logic, we analyze a representative case from the SQL-MCTS dataset as shown in Figure~\ref{fig:case}. The original query suffers from significant computational redundancy because it repeats a complex aggregation involving several tables. This heavy operation is executed once in the main query to retrieve category sales and again within the subquery to calculate the global average threshold, forcing the database to scan the large fact tables twice. Existing rule-based and retrieval-augmented approaches, such as R-Bot~\cite{r-bot}, struggle to resolve this inefficiency. This is because they depend on predefined transformation rules, limiting their ability to identify duplicated computation patterns dispersed across distant query blocks. In contrast, LASER-14B successfully identifies this inefficiency and reconstructs the query using Common Subexpression Elimination. It extracts the repeated logic into a CTE~\cite{cte}, which is computed once and then reused to derive the average value in a subsequent step. This logical refactoring effectively halves the heavy lifting, reducing the execution time from over 30 seconds to approximately 10 seconds.
Moreover, our SQL-MCTS dataset contains many such challenging query structures. This diversity and complexity present a significant advantage for training models.

\subsection{Deployment at ByteDance}
To validate the practical applicability of our approach, we deployed the LASER-14B model within ByteDance's production environment.
We evaluated its effectiveness using several real-world datasets derived from diverse business scenarios, such as E-commerce, Finance, and Video Streaming. This collection encompasses 129 databases and contains approximately 542GB of data. The evaluation workload consisted of 462 actual queries.
From this workload, LASER-14B identified and rewrote 69 queries that exhibited potential for optimization. Empirical results demonstrate that 73\% of these rewritten queries achieved performance gains, yielding an average latency reduction of 23.1\% compared to the original execution times. These findings robustly verify the effectiveness and reliability of the LASER framework in handling real-world business scenarios.

\section{Conclusion}
\label{sec:conclusion}

In this paper, we introduced LASER, a data-centric approach for low-cost and efficient SQL query rewriting based on SQL-GRPO. 
By leveraging slow query generation to create high-quality training data, we synthesized over 11,000 complex slow queries, called SQL-MCTS. We then utilized Group Relative Policy Optimization, combined with Complexity-Adaptive Dynamic Rollout and Anchored Group Advantage, to enhance the performance of small models for SQL query rewriting. We demonstrate the effectiveness of LASER by training Qwen3-8B and Qwen3-14B models, and testing them on several public benchmarks. 
Extensive experiments show that our approach significantly improves model performance, confirming LASER's ability to enhance the efficiency and accuracy of small models in real-world query optimization scenarios.

\balance
\bibliographystyle{ACM-Reference-Format}
\bibliography{sample}

\end{document}